# Room-Temperature Optical Spin Polarization of an Electron Spin Qudit in a Vanadyl – Free Base Porphyrin Dimer

Alberto Privitera,[1,2*] Alessandro Chiesa,[3] Fabio Santanni,[4] Angelo Carella,[1,5] Davide Ranieri,[4] Andrea Caneschi,[2] Matthew D. Krzyaniak,[1] Ryan M. Young,[1] Michael R. Wasielewski,[1*] Stefano Carretta,[3*] and Roberta Sessoli[4*]

[1] Department of Chemistry, Center for Molecular Quantum Transduction, and Paula M. Trienens Institute for Sustainability and Energy, Northwestern University, 60208-3113, Evanston, IL (USA)
[2] Department of Industrial Engineering, University of Florence & UdR INSTM Firenze, 50139, Firenze, Italy
[3] Department of Mathematical, Physical and Computer Sciences, University of Parma & UdR INSTM, 43124, Parma, Italy.
[4] Department of Chemistry "U. Schiff", University of Florence & UdR INSTM Firenze, 50019, Sesto Fiorentino, Italy
[5] Department of Chemical Sciences, University of Padova, 35134, Padua, Italy

**ABSTRACT:** Photoexcited organic chromophores appended to molecular qubits can serve as a source of spin initialization or multilevel qudit generation for quantum information applications. So far, this approach has been primarily investigated in chromophore/stable radical systems. Here, we extend this concept to a meso-meso linked oxovanadium(IV) porphyrin – free base porphyrin dimer. Femtosecond transient absorption experiments reveal that photoexcitation of the free base porphyrin leads to picosecond triplet state formation via enhanced intersystem crossing. Time-resolved electron paramagnetic resonance (TREPR) experiments carried out at both 85 K and room temperature reveal the formation of a long-lived spin-polarized quartet state through triplet–doublet spin mixing. Notably, a distinct hyperfine structure arising from the interaction between the electron spin quartet state and the vanadyl nucleus ($^{51}$V, I=7/2) is evident, with the quartet state exhibiting long-lived spin polarization even at room temperature. Theoretical simulations of the TREPR spectra confirm the photogenerated quartet state and provide insights into the non-Boltzmann spin populations. Exploiting this phenomenon affords the possibility of using photoinduced triplet states in porphyrins for quantum information as a resource to polarize and magnetically couple molecular electronic or nuclear spin qubits and qudits.

## Introduction

Quantum bits, or qubits, are the fundamental building blocks of Quantum Information Science (QIS).[1-2] Traditionally, qubits have been engineered using solid-state systems such as superconducting circuits,[3] trapped ions,[4] quantum dots,[5] nitrogen-vacancies (NV) defects in diamonds,[6] and semiconductor dopants,[7] employing a top-down strategy. In contrast to these conventional methodologies, chemical synthesis provides a unique pathway for creating novel molecular qubits, taking full advantage of the quantum properties of matter at the atomic scale.[8-9] Molecular qubits offer several advantages, including tunable properties through synthetic flexibility, long coherence times, surface processability, and the potential to form extended qubit arrays.[8-9] Moreover, their multilevel structure enables their exploitation as qudits.[10] Despite these strengths, a critical challenge remains: molecular systems typically rely on thermally polarized electron spins with well-defined initial spin states only available at high magnetic fields and sub-kelvin temperatures.[11-12] A promising approach to overcome this limitation draws inspiration from the NV centers, which can be optically pumped to achieve highly spin-polarized states and read out using optically detected magnetic resonance (ODMR) spectroscopy.[13-14] To achieve similar optical pumping and addressability properties in molecular qubits, researchers have recently focused on photoexcited organic chromophores (C) covalently bound to spin-doublet qubits (QB, $S$=1/2).[15-18] Figure 1 depicts a typical photophysical pathway of a C-QB system.

Following photoexcitation of the chromophore, the doublet ground state of the compound ($^2$[$^1$C–$^2$QB], **D₀**) is optically pumped to its first excited state, also called sing-doublet state ($^2$[$^{1*}$C–$^2$QB], **D₁**). From this state, enhanced intersystem crossing (EISC) can occur driven by the exchange interaction between the two spin-paired electrons on $^{1*}$C with the unpaired electron on $^2$QB to generate [$^{3*}$C-$^2$QB].[19-21] In the strong exchange coupling regime ($J \gg \Delta g \mu_B B$), where $J$ is the magnetic exchange interaction between the triplet $^{3*}$C and the doublet $^2$QB and $\Delta g$ the difference in their $g$-factors, the resulting three unpaired electron system is best described at high magnetic fields as a two-state trip-doublet ($^2$[$^{3*}$C–$^2$QB], **D₂**) and a four-state trip-quartet ($^4$[$^{3*}$C–$^2$QB], **Q**) separated by the exchange interaction.[18, 22] Since **D₁** and **D₂** have the same spin multiplicity, the transition from $D_1$ to $D_2$ is more rapid than that to **Q**. Furthermore, decay of **Q** to the ground doublet state $D_0$ is spin forbidden, allowing a sufficiently long lifetime to probe and manipulate **Q** using resonant microwave pulses. Importantly, all the processes discussed so far are spin selective, leading to the rise of spin polarization, i.e., the population of the spin sublevels is out of thermal equilibrium. This mechanism potentially eliminates the need for millikelvin temperatures to initialize a molecular qubit. Various spin polarization mechanisms may occur in a photoexcited C-QB system, depending on the coupling regime and specific magnetic interactions involved in spin dynamics.[15, 19, 23-24]

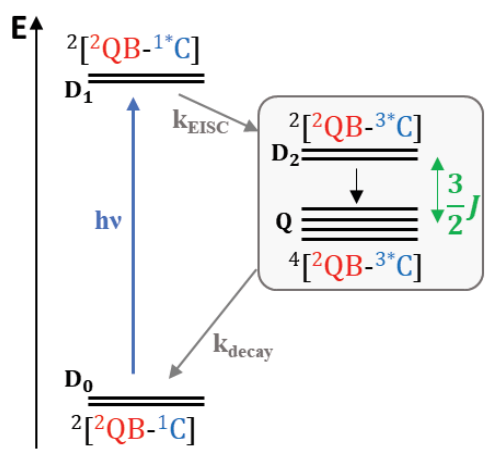

**Figure 1.** Schematic representation of the photophysical pathways in a photoexcited chromophore (C) - spin-doublet qubit (QB) system. Detailed photophysical pathways are introduced in the text.

So far, the study of these processes for QIS applications has primarily targeted organic chromophore-radical systems.[15] These systems serve as excellent models due to their favorable coherence properties, simple photophysics, and narrow EPR linewidths that facilitate microwave addressability.[15] However, comparatively little attention has been given to extending this concept to molecular qubits based on metal ions.[25-26] Among this qubit family, metalloporphyrins have emerged as prominent candidates.[26] In particular, vanadyl ($V^{IV}O$) porphyrins represent key model systems thanks to their excellent coherence properties and their multilevel nature arising from the electron and nuclear spin states of vanadium ($S=1/2$ and 99.75 % $^{51}V$ with $I=7/2$).[27-32] The latter property is noteworthy as vanadyl metalloporphyrins possess the potential to serve as multilevel elementary units that enhance quantum logic capabilities and facilitate quantum error correction at the molecular level.[33-36] In addition, the vanadyl unit is crucial in facilitating significant magnetic exchange interactions when connected to other porphyrin units while maintaining good coherence properties and single-unit addressability, which is key for implementing quantum gates.[37-39] Finally, due to their structural stability, vanadyl porphyrin qubits can be thermally evaporated onto surfaces to form monolayers and are promising for the development of solid-state QIS architectures.[40-42]

In view of this, exploiting chromophore-qubit photoinduced interactions in VO porphyrins would represent a significant leap forward in QIS research. To our knowledge, the only prior spin photophysical studies on VO porphyrins focused on monomers and revealed the photogeneration of a quartet state with absorptive out-of-equilibrium spin polarization at low temperatures.[43-44] Decoupling the chromophore unit from the VO qubit would enable independent optimization of each component and precise tuning of their exchange interactions and spin polarization pathways. In particular, this approach allows maintaining the isolated qubit properties while utilizing the chromophore to control its state or couple it with other logical units. Here, we present the investigation of the photophysical and spin polarization evolution of the vanadyl porphyrin – free base porphyrin *meso-meso* linked dimer, namely [oxo(10,20-diphenylporphyrinato-5-yl)vanadium(IV)]-10,20-diphenylporphyrin, **VO-FP** hereafter (see Figure 2a). The individual building blocks, [oxo(5,10,15-triphenylporphyrinato)vanadium(IV)], **VO**, and 5,10,15-triphenylporphyrin, **FP**, are also investigated for comparison. Transient absorption (TA) experiments demonstrate that upon photoexcitation of **FP**, the triplet state formation occurs within picoseconds, promoted by EISC. Time-resolved electron paramagnetic resonance (TREPR) spectroscopy shows a totally emissive signal at 85 K, demonstrating a largely out-of-equilibrium polarization. Moreover, even at room temperature TREPR displays clear features of the quartet state and the $^{51}V$ hyperfine structure resulting from the ferromagnetic coupling of the excited triplet state on the chromophore with the unpaired electron on the $VO^{2+}$ center. Simulations of spectra recorded orienting the molecules in liquid crystals reveal significant non-Boltzmann spin population across both electronic and nuclear spin sublevels. The study lays the groundwork to utilize the free-base porphyrin unit coupled with vanadyl porphyrin as a benchmark system for harnessing the diverse spin properties of metal-based molecular qubits in light-driven QIS applications.

**Experimental section**

**Sample preparation. VO-FP** was prepared following the procedure we previously developed.[37] We selected triphenylporphyrin as the basis for preparing the individual building blocks, **VO** and **FP**, to ensure that each compound has only a single unsubstituted *meso* position.[38] Further details and characterization are given in SI (Figures S1-S5). Room-temperature femtosecond and nanosecond transient absorption (fsTA and nsTA) optical experiments were performed in toluene solutions, which were prepared in 2 mm path length glass cuvettes and degassed with three freeze-pump-thaw cycles. Low temperature experiments were performed in glassy butyronitrile (PrCN) at 85 K or toluene solution at 125 K. The butyronitrile was distilled, deoxygenated, and stored in the glove box in which the sample solutions were prepared. The low temperature sample cell, composed of two quartz windows separated by a Teflon spacer (~2 mm), was filled with the solution, and assembled under inert gas atmosphere in the glove box. The sample concentration was adjusted to have an optical density (O.D.) of about 0.3–0.6 at the excitation wavelength in the sample cuvette. For the measurements at a 640 nm excitation wavelength, O.D. was maintained at approximately 0.1 to prevent aggregation.

For all EPR measurements, the samples were prepared with concentrations similar to those used in the TA experiments. Solutions (~50 μL) were loaded into quartz tubes (2.40 mm outer diameter, 2.00 mm inner diameter), subjected to three freeze-pump-thaw cycles on a vacuum line ($10^{-4}$ Torr), and sealed with a hydrogen torch. For low temperature experiments, the samples were pre-frozen in liquid nitrogen before being inserted into the pre-cooled resonator at 85 K.

**Optical Spectroscopy.** Steady-state absorption spectra were acquired on a Shimadzu 1800 spectrophotometer. The fsTA and nsTA experiments were conducted using a previously described instrument.[45-46] Both measurements were performed by using a regeneratively amplified Ti:sapphire laser system operating at 1 kHz repetition rate to generate 828 nm pulses, which create the ca. 545 nm or 640 nm excitation pulses using a commercial collinear optical parametric amplifier (TOPAS-Prime, Light Conversion, LLC). Low-temperature fsTA and nsTA experiments were performed in a Janis VNF-100 cryostat (Lakeshore Cryotronics) using a Cryo-Con 32B (Cryogenics Control Systems, Inc.) temperature controller. All TA spectra were acquired by using an excitation energy of about 1 μJ/pulse. The data were background-subtracted and chirp-corrected using a lab-written MATLAB program. The TA data were subjected to global kinetic analysis to obtain the evolution-associated and kinetic parameters as described in detail previously.[47]

**Time-resolved continuous-wave EPR (TREPR) spectroscopy.** All EPR measurements at X-band were made on a Bruker Elexsys E580 EPR spectrometer equipped with a 3 mm split ring resonator (ER4118X-MS3) ($\nu_{MW}$ ~ 9.5 GHz, MW power = 0.2 mW). The temperature was controlled by an Oxford Instruments CF935 continuous flow optical cryostat using liquid nitrogen. For transient CW EPR studies, the sample was photoexcited at 550 and 640 nm with 7 ns pulses generated via an optical parametric oscillator (Spectra-Physics BasiScan) pumped with the 355 nm output of a frequency-tripled Nd:YAG laser (Spectra-Physics Quanta-Ray Lab-150 or Lab-170-10H) operating at a repetition rate of 10 Hz. The laser light was coupled into the resonator with a fiber optic and collimator placed outside the cryostat window, resulting in about 1 mJ/pulse passing into the resonator and exciting the sample. Following photoexcitation, the transient magnetization time traces were acquired as a function of the magnetic field using direct diode detection under continuous microwave irradiation. The data were processed by first subtracting the background signal prior to the laser pulse for each kinetic trace (at a given magnetic field) and then subtracting the signal at off-resonance magnetic fields for each spectrum (at any given time).

**Spectral simulations.** EPR simulations were performed using a home-built code developed in the Matlab scripting environment, supplemented by routines from the EasySpin simulation package.[48] Further details of the spectral simulations are provided in the text.

## Results and discussion

**The molecular qubit-chromophore system.** An effective molecular qubit-chromophore system should meet specific optical and spin properties. Primarily, the molecular qubit unit should exhibit favorable ground-state magnetic properties, such as long coherence times and a well-defined EPR spectrum. Individually addressable nuclear transitions in a multilevel hyperfine structure represent an additional resource in QIS.[10, 31] Additionally, the chromophore should possess strong absorption and potential for selective excitation. In the molecular qubit-chromophore system, intersystem crossing (ISC) should be favored over alternative detrimental photophysical pathways, such as non-radiative internal conversion, energy transfer, or electron transfer.[26] We selected **VO-FP** as an ideal candidate (Figure 2a) for free porphyrin selective optical excitation.[49] The following ISC generates a triplet state that remains localized on the chromophore unit.[49-51]

The steady-state UV-vis absorption spectra of the three molecules in toluene solution at room temperature are shown in Figure 2b. As typically observed for porphyrin derivatives,[52] the electronic absorption spectrum consists of two distinct sets of bands, the Soret bands in the near-UV region and the 'Q-bands' in the visible region, both of which involve ππ* excitation of the macrocycle. In **VO-FP**, the intense Soret band in the 350-500 nm region exhibits splitting due to significant excitonic coupling between the individual porphyrin units.[53] In contrast, the Q-bands in the 500-700 nm region show only slight broadening and changes in relative intensities. Notably, the absorption peak of **FP** at 640 nm, where **VO** does not absorb, remains unchanged in both the monomer and the dimer, allowing for the selective excitation of free-base porphyrin in the dimer.

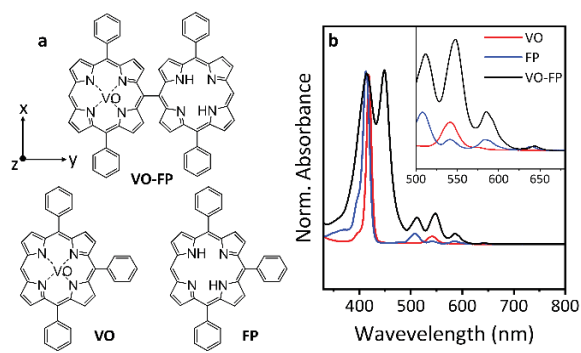

**Figure 2.** Molecular structures and steady-state UV/vis absorption spectra of the three investigated molecules: **VO** = vanadyl monomer, **FP** = free-base monomer, and **VO-FP** = vanadyl-free base dimer. The system of reference for the VO unit is characterized by $y$ along the meso-meso bond direction, $z$ perpendicular to the plane of VO, and $x$ perpendicular to both $y$ and $z$.

**Transient absorption spectroscopy.** The fs/nsTA spectra of the **VO** monomer recorded in toluene solution at room temperature after photoexcitation at 550 nm are shown in Figure S6. Immediately after photoexcitation, the typical intense Soret ground-state bleaching (GSB) signal at ∼ 400 nm and the stimulated emission (SE) of the Q(0,1) band at ∼ 650 nm overlap with the excited state absorption (ESA) covering almost the entire visible range and part of NIR region.[43, 54-56] The SE at ∼ 650 nm decays in less than 300 fs, suggesting the presence of an enhanced intersystem crossing (EISC).[43, 54-56] The GSB and the broad ESA typical of the porphyrin triplet state persist over hundreds of nanoseconds before entirely decaying to the ground state. The time constants of the excited state deactivation were determined by performing a global kinetic analysis of the fsTA data. The results, including the decay- and evolution-associated spectra (EAS), are presented in the SI. The analysis revealed that the **VO** excited singlet state decays into a triplet state with a time constant faster than 300 fs, which is below the time-resolution of our TA apparatus. The presence of the VO paramagnetic center promotes EISC through exchange coupling interaction, making it two orders of magnitude faster compared to the free-base porphyrin (∼ 12 ns).[45,57] This effect has been observed in other paramagnetic metalloporphyrins in previous literature.[45-46, 49-50] It is important to note that TA data do not distinguish between the nearly degenerate trip-doublet (**D**$_2$) and trip-quartet (**Q**) states resulting from the interaction of the triplet state on the porphyrin ligand with the $S$=1/2 of vanadyl. The porphyrin triplet state then decays to the ground state with a time constant of 74 ± 3 ns.

A similar photophysical landscape, but with different time constants, is observed in the **VO-FP** dimer. In Figure 3, the fs/ns TA spectrum of **VO-FP** excited at 640 nm in toluene at room temperature is shown. At 640 nm, excitation selectively targets the free-base unit. Extended transient absorption (TA) characterization was performed in toluene at room temperature and 125 K, and in butyronitrile at 85 K (Figures S7-S10), following laser excitation at 640 nm and 545 nm for comparison. Global analysis of the spectra revealed no significant differences in the photophysical pathways of **VO-FP** across different solvents and wavelengths (further details in the SI). Global kinetic analysis of the data provides the EAS with the indicated time constants for the formation and decay of the different species. The initial signal, characterized by the GSB of the two split Soret bands, the four Q bands, and the broad ESA extending in the visible and NIR regions, is assigned to the singlet excited state of **VO-FP**. The SE at ~645 nm is not visible when excited at 640 nm due to pump scatter, but it is observable when excited at 545 nm (Figures S8, S10). After 1.1 ± 0.3 ps (2.90 ± 0.05 ps at 85 K, Figure S7), the excited singlet state decays into a long-lived triplet state. In Figures S7-10, the characteristic broad absorption extending from 600 nm through the NIR region is also visible. The slower EISC rate in **VO-FP** compared to **VO** can be attributed to the larger separation between the singlet excited state and the paramagnetic vanadyl ion, thus confirming that in **VO-FP** the excitation is spatially confined on the free-base.[49] Finally, the triplet state decays to the ground state in 46.3 ± 0.4 μs (160 ± 1 μs at 85 K, Figure S7). This decay is significantly slower than in **VO** but faster than in free-base porphyrins, where the triplet state typically lasts on the order of milliseconds (at 77 K).[57]

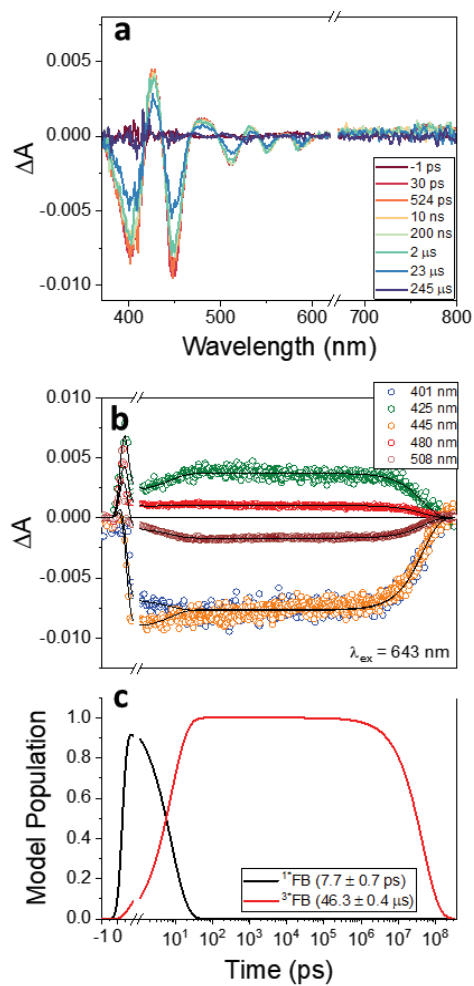

**Figure 3**. (a) Room-temperature fs/nsTA spectra of **VO-FP** in toluene excited at 640 nm and recorded at selected delay times. (b) Selected wavelength kinetic fits and (c) population dynamics obtained by globally fitting the fs/nsTA data. The mechanism assumed to fit the data is A($^2$VO-$^1$*FP) →B($^2$VO-$^3$*FP)→GS. At short times, the solvent response due to the high laser pump fluence (2 μJ/pulse) was considered in the fitting but is not shown.

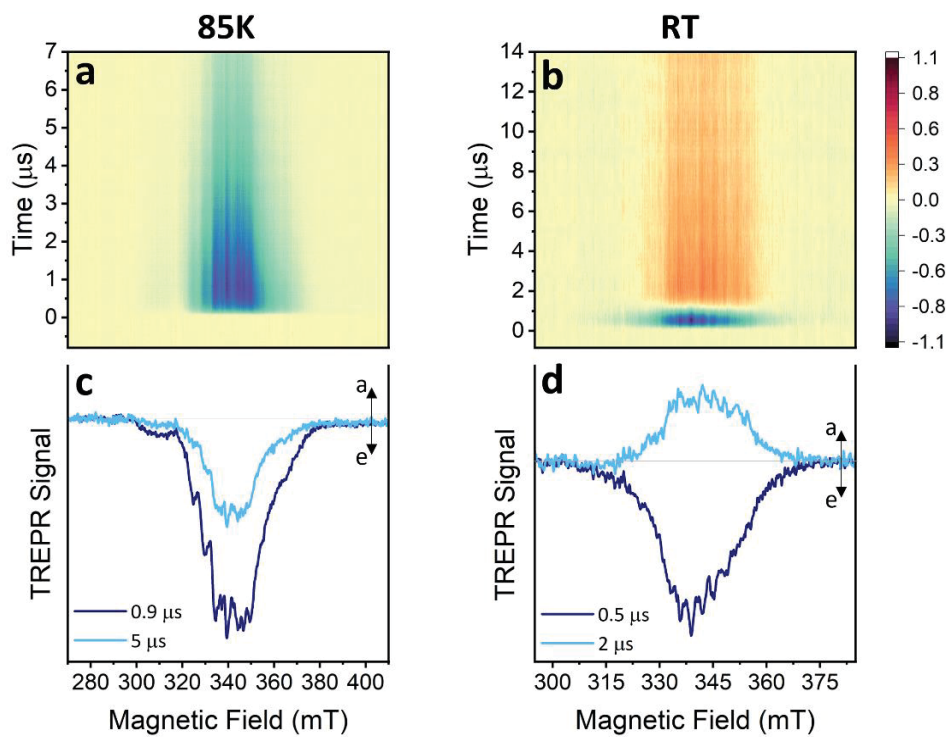

**Figure 4**. (a,b) Normalized 2D experimental TREPR contour plots of **VO-FP** in toluene acquired at 85 K and RT after a 650 nm laser pulse (7 ns, 2 mJ). Color legend: red = enhanced absorption, blue = emission, yellow = baseline. (c,d) Normalized 1D experimental TREPR spectra taken at representative times after the laser pulse (integrated time window = 100 ns). Arrows legend: a = enhanced absorption, e = emission.

**TREPR spectroscopy.** EPR experiments were performed to characterize the magnetic properties of the photoexcited spin states. The dark-state EPR characterization, reported in our previous publications,[37-38] shows no difference in the **g**- and **A**-tensors of vanadyl between **VO** and **VO-FP**. The TREPR spectra of the **FP** and **VO** in frozen toluene solution acquired at $T$ = 85 K after 550 nm laser excitation are shown in Figures S11 and S12, respectively. **FP** exhibits the characteristic triplet spectrum of free-base porphyrins, simulated with parameters listed in Table S1, consistent with previous literature findings.[58] Conversely, **VO** displays a broad spectrum spanning from 260 mT to 420 mT, exhibiting solely positive intensity. As the spectra were recorded in direct detection, positive signals correspond to absorptive (a) transitions, while negative signals correspond to emissive (e) ones. This spectrum aligns well with prior observations on 2,3,7,8,12,13,17,18-octaethylporphynato oxovanadium(IV)[43] and can be attributed to the long-lived trip-quartet state of **VO**. No significant TREPR signal was observed at room temperature, consistently with the fast decay of the triplet state.

The TREPR spectrum of **VO-FP** in frozen toluene solution at 85K is shown in Figure 4a. The 1D spectra taken at 0.9 μs and 5 μs after a 640 nm laser pulse are shown in Figure 4c. Notably, the comparison between spectra acquired with 640 nm and 550 nm excitation (Figure S13) reveals no significant differences in the polarization pattern, confirming the TA results. The spectra exhibit an intense net emissive polarization from 300 mT to 380 mT, which persists beyond 7 μs. This behavior is different from what we observed for **VO**, which shows only absorptive spin polarization. The decay of the spin polarization is governed by the spin-lattice relaxation time, while TA indicates the excited state lasts for hundreds of microseconds. Notably, a clear hyperfine structure attributable to the $I$=7/2 vanadyl nucleus is visible; the hyperfine coupling to the four nitrogen nuclei is not resolved, similar to what is observed in the ground-state doublet.

Taking advantage of the anisotropy of the **g**- and **A**-tensors of vanadyl porphyrin (Table 1), we obtained detailed information from the orientation-dependent spectra of **VO-FP** dissolved in the nematic liquid crystal 4-cyano-4'-(n-pentyl)biphenyl (5CB). The solution in 5CB was aligned in a magnetic field at 295 K, then rapidly frozen to 85 K, aligning the long axes of 5CB molecules along the magnetic field.[59] The aligned and unaligned (isotropic) TREPR spectra of **VO-FP** in 5CB were obtained by photoexciting the samples with a 640-nm laser pulse and are shown in Figure 5 for a delay of 2 μs, while the complete time dependence survey is available in Figure S14. The unaligned spectrum shows no significant difference with respect to the spectrum in toluene (Figures 4c/5e). The aligned spectra were taken at two different sample orientations, referred to as parallel (5CB long axis parallel to $B_0$) and perpendicular (5CB long axis perpendicular to $B_0$). To a first approximation, **VO-FP** aligns with the *meso-meso* bond axis (Figure 2a) parallel to the 5CB.[59] Thus, in the parallel orientation, the $g_{VO,y}/A_{VO,y}$ transitions of the vanadyl are preferentially probed, while in the perpendicular orientation the $g_{VO,z}/A_{VO,z}$ and $g_{VO,x}/A_{VO,x}$ ones. This accounts for the larger and clearer hyperfine pattern observed in the perpendicular spectrum. Further details are provided in the spectral simulation section.

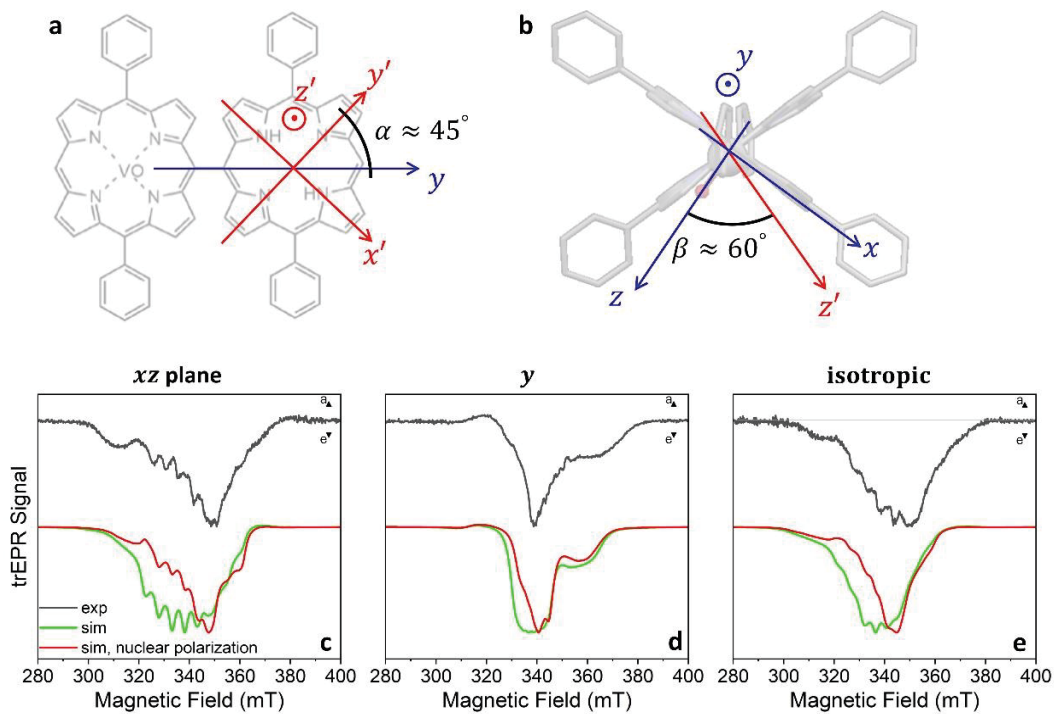

**Figure 5**. (a,b) Orientation of the principal axes of the VO and FP units. (c-e) Normalized 1D experimental TREPR spectra (black line) and spectral simulations (green and red lines) of VO-FP aligned in the nematic liquid crystal 5CB at 85 K, taken 2 μs after a 640-nm laser pulse (duration 7 ns, energy 2 mJ). The **VO-FP** molecules are oriented with their long axis (y) perpendicular (c) and parallel (d) relative to the external magnetic field direction, while they are in an isotropic frozen solution in (e). Simulations assume a Lorentzian line width (1.8 mT) and a Gaussian distribution of orientations with a standard deviation of 10°, using parameters listed in Table 1. Simulation legend: green line = simulation with electronic spin polarization only; red line = simulation including both electronic and nuclear spin polarization.

The assignment of the TREPR spectra to a specific state is not straightforward since they could contain contributions from the trip-doublet **D$_2$**, trip-quartet **Q**, and/or the ground state **D$_0$** depending on the nature of the excited-state dynamics and the values of the magnetic parameters. In the strong exchange regime, the contribution of the $^{51}$V hyperfine coupling for the trip-quartet and trip-doublet states is + 1/3 and -1/3 of the value in the ground state, respectively.[60] The observed hyperfine pattern in Figure 5c, where the z axis of the vanadyl is sampled, provides a hyperfine spacing of approximately 5 mT (~ 140 MHz), about one third of the vanadyl value, ruling out the observation of the ground doublet. Additionally, previous literature suggests that photoexcited paramagnetic species observed via TREPR in metalloporphyrins are typically associated with the longer-lived trip-quartet state rather than the rapidly decaying trip-doublet state.[43, 61]

The potential application of molecular qubits at room temperature prompted us to investigate whether spin polarization survives in toluene solution at room temperature. The spectra are significantly simplified, even if the tumbling of **VO-FP** at room temperature does not completely average out the anisotropies, as shown in the dark-state EPR spectra in Figure S15. The TREPR spectrum of **VO-FP** in toluene, using 640 nm laser pulses, is shown in Figure 4b, along with the 1D spectra taken at representative times in Figure 4d. Shortly after the laser pulse (0.5 μs), the spectrum exhibits a net emissive spin-polarization extending from 315 mT to 365 mT. Over the course of 1 μs, this spin-polarization evolves due to spin-lattice relaxation, becoming purely absorptive and persisting for longer than 14 μs. Notably, the characteristic pattern with reduced hyperfine spacing corroborates that we are detecting the signal resulting from the coupling of the vanadyl qubit with the excited chromophore triplet.

**Spectral simulations.** We describe the system in the photoexcited state by the general two-spin Hamiltonian:

**Table 1** Parameters of the simulations reported in Figure 5. Left column: Fitting parameters related to the out-of-equilibrium populations of the density matrix within the trip-quartet used to reproduce experimental spectra. The parameters a and r determine the traceless diagonal part of the density matrix on the four electronic spin sublevels (bottom line) along different orientations of the external field with respect to the molecular axes. $\rho_N$ indicates the normalized populations of the nuclear spin sublevels. Right: Hamiltonian parameters, kept fixed from previous studies for VO or independent EPR measurements on FP.

| FITTING PARAMETERS (POPULATIONS) | | FIXED PARAMETERS (HAMILTONIAN) | |
|---|---|---|---|
| **a** | $(0.11, -0.002, -0.027)$ | $g_{FP}$ | $2.0023$ |
| **r** | $(0, -0.01, 0)$ | $D, E$ | $(1135, 235)$ MHz |
| $\rho_N$ | $(0.146, 0.078, 0.194, 0.126, 0.117, 0.165, 0.078, 0.097)$ | $\mathbf{g}_{VO}$ | $(1.985, 1.985, 1.964)$ |
| | | $\mathbf{A}_{VO}$ | $(162, 162, 475)$ MHz |
| $\rho_S^x, \rho_S^y, \rho_S^z$ | $(-0.101, -0.025, 0.078, 0.047), (-0.059, -0.066, 0.037, 0.089), (0.012, -0.012, -0.012, 0.012)$ | | |

$$H = J\mathbf{s}_{FP} \cdot \mathbf{s}_{VO} + d\left(s_{FP,x}s_{VO,x} + s_{FP,z}s_{VO,z} - 2s_{FP,y}s_{VO,y}\right)$$
$$+ D\left[s_{FP,z'}^2 - \frac{s_{FP}(s_{FP}+1)}{3}\right]$$
$$+ E\left(s_{FP,x'}^2 - s_{FP,y'}^2\right) + g_{FP}\mu_B \mathbf{B} \cdot \mathbf{s}_{FP}$$
$$+ g_{VO,\perp}\mu_B\left(B_x\, s_{VO,x} + B_y\, s_{VO,y}\right)$$
$$+ g_{VO,z}\mu_B\, B_z\, s_{VO,z}$$
$$+ A_{VO,\perp}\left(I_{VO,x}\, s_{VO,x} + I_{VO,y}\, s_{VO,y}\right)$$
$$+ A_{VO,z}\, I_{VO,z}\, s_{VO,z}$$

where $s_{FP} = 1, s_{VO} = 1/2$ are the spin of the free-base porphyrin (after ISC) and of the VO qubit, respectively. The first two terms describe the exchange ($J$) and dipole-dipole ($d$) contributions. The following two terms model the axial and rhombic zero-field splitting anisotropy of the FP triplet (parameterized by $D$ and $E$). The next are the Zeeman interactions with the external field $\mathbf{B}$ of FP (isotropic, with spectroscopic factor $g_{FP}$) and VO (characterized by transverse $g_{VO,\perp}$ and longitudinal $g_{VO,z}$ components of the spectroscopic tensor $\mathbf{g}_{VO}$). The last two terms represent the hyperfine coupling of $^{51}$VO $s_{VO} = 1/2$ with its $I_{VO} = 7/2$ nuclear spin, characterized by the axial hyperfine tensor $\mathbf{A}_{VO}$ with transverse and longitudinal components $A_{VO,\perp}$ and $A_{VO,z}$, respectively. All these terms are referred to the local principal axes sketched in Figure 5. In particular, $xyz$ and $x'y'z'$ indicate the principal axes of the VO doublet and of the FP triplet, respectively. Since $g_{FP}$ is isotropic, the principal axes of the dipolar spin-spin interaction are aligned so that its longitudinal component $z^{dip}$ axis is along the bond direction, i.e. parallel to $y$. Importantly, $x'$ and $y'$ are rotated of about $\alpha \approx 45°$ with respect to $y$, as indicated in Figure 5a.[62-63] The $z$ and $z'$ axes are orthogonal to the respective porphyrin planes, and hence the angle between $z$ and $z'$ ($\beta$ in Figure 5b) corresponds to the dihedral angle between the two porphyrin planes, $\approx 60°$.[37-38] The relative orientation of these tensors is kept fixed in the simulations. The VO **g**- and **A**-tensors are known from previous studies,[37-38] while the g-factor and zero-field splitting tensor of the FP were fitted from EPR data on the isolated monomer **FP** (see Figure S11). Finally, the dipole-dipole interaction $d$ is computed by assuming two spin centers with known distance of 0.84 nm, resulting in $d \approx 90$ MHz. Hence, the only fitting parameters are the exchange coupling $J$ and the non-Boltzmann population of the eigenstates.

The Hamiltonian above is general and allows exploring different parameter regimes. In the strong exchange limit ($J \gg |g_{FP} - g_{VO}|\mu_B B, D, E, A$), the eigenstates of the photoexcited system are organized into two total spin multiplets (i.e. trip-doublet and trip-quartet), separated by an energy gap $3/2\, J$. Once this hierarchy of parameters is established, the precise value of $J$ does not affect our conclusions. Hence, we assumed $J = 1$ cm$^{-1}$, a reasonably smaller value compared to the literature,[44] due to the larger separation and torsion angle between the two porphyrin planes. At the same time, the assumed exchange interaction is two orders of magnitude larger than that between the two VO units in the VO-VO *meso-meso* linked porphyrin dimer, where very little spin density is delocalized on the porphyrin rings.[38] Furthermore, the long lifetime of the photoexcited state suggests a ferromagnetic exchange interaction, with the trip-quartet lower in energy than the trip-doublet (as sketched in Figure 1). Spin polarization within the trip-quartet can then be accumulated by ISC from the trip-doublet state.[44]

The non-Boltzmann population of the trip-quartet can be expressed as the traceless diagonal part of the reduced density matrix within the quartet subspace. Following Kandrashkin et al.,[44] we express it in powers of $S_z$ operators as follows:

$$\rho_S(\theta, \phi) = a_2(1 - 3\cos^2\theta)\left[S_{Q,z}^2 - \frac{S_Q(S_Q+1)}{3}\right]$$
$$+ r_2 \sin^2\theta \cos 2\phi \left[S_{Q,z}^2 - \frac{S_Q(S_Q+1)}{3}\right]$$
$$+ a_1 \sin^2\theta\, S_{Q,z} + r_1 \cos^2\theta\, S_{Q,z}$$
$$+ a_3 \sin^2\theta\, S_{Q,z}^3 + r_3 \cos^2\theta\, S_{Q,z}^3$$

where Q denotes the quartet state, the angles represent the orientation of the molecular frame relative to the external field, with the $z_Q$ axis orthogonal to the plane of FP and the $x_Q$ axis resulting from the combination of zero-field splitting and dipole-dipole tensors (see Figure 5a). This expression accounts for both axial and non-axial contributions to the polarization (arising from different components of spin-orbit coupling in the ISC from the trip-doublet).

This treatment allows us to model the spin polarization of the trip-quartet, $\rho_S$, using at most 6 fit parameters $a_i, r_i$ independently from the orientation of the molecule with respect to the external field. Hence, we perform a simultaneous fit of the experimental spectra taken at 85 K along the parallel ($y$) and perpendicular ($xz$) orientations and in the isotropic frozen solution (Figure 5c-e, green lines). The parameters **a** and **r** used in the simulations are listed in Table 1 (left), along with the fixed parameters of the Hamiltonian (right, known from previous studies). The resulting population of the quartet states along different directions (defined apart from an arbitrary multiplicative constant factor) are also shown in Table 1 in order of ascending energy in the eigenbasis. We highlight that these populations are largely out of thermal equilibrium. Indeed, the leading term in $\rho_S^{x,y}$ is proportional to $+S_z$, a situation opposite to the thermal equilibrium one, in which $\rho_S$ is approximately $\propto -S_z$.

The simulations give rise to the observed emissive character of the spectrum. In addition, the overall width is correctly reproduced, consistently with the projection of the VO hyperfine coupling within the quartet subspace ($A_Q = \frac{A}{3}$).[44] This rules out a significant polarization of the ground state **D₀**, which would lead to a broader spectrum. However, the relative intensities of the different nuclear spin sub-levels are not accurately reproduced. To address this discrepancy, we include different populations of the hyperfine states within the quartet as adjustable parameters, i.e.

$\rho_0 = \rho_S(\theta, \phi) \otimes \rho_N$.

This is a reasonable assumption since stationary populations depend on the different transition rates between **D₂** and **Q** sublevels. These rates depend not only on the different $m_S$ involved in each transition, but also on the hyperfine sublevels $m_I$. As a result, we obtain a good simulation of the spectra by assuming a sizable polarization of the hyperfine states, far from thermal equilibrium (see Table 1).[64]

**Conclusions**

In this study, we explored the optical and magnetic properties of a photogenerated molecular three-spin system composed of a FP chromophore triplet and a VO qubit. Upon selective photoexcitation of FP, the excited singlet state of FP is rapidly quenched by exchange-mediated EISC, leading to the formation of the FP excited triplet state. Using the combination of TREPR experiments and theoretical simulations, we identified the generation of a transient species with quartet multiplicity, indicative of ferromagnetic interaction between the FP triplet and the VO doublet within the strong exchange coupling regime ($J \gg |g_{FP} - g_{VO}|\mu_B B, D, E, A$). Compared to VO monomers, both from our experiments and literature reports,[43-44] the **VO-FP** trip-quartet state exhibits significantly extended lifetimes, likely due to reduced dipole-dipole interactions between the spins. This extended lifetime allows the observation of spin-polarized TREPR signals even at room temperature.

Notably, the **VO-FP** system displays a unique emissive non-Boltzmann spin population of the trip-quartet state that persists for several microseconds at room temperature, reflecting a long spin-lattice relaxation time ($T_1$). This extended $T_1$ makes the system an excellent benchmark for investigating spin-polarization mechanisms in metal-based molecular qubits. The decoupling of the FP chromophore from the VO qubit allows for the independent optimization of each component, enabling precise tuning of spin-selective processes across varying exchange coupling regimes (from strong to moderate and weak) through ligand engineering. This flexibility opens up the potential to observe spin polarization in the **D₀** ground state, providing a promising avenue for qubit initialization in pseudo-pure spin states.[17]

Importantly, the ISC mechanism amplifies the population difference between specific pairs of nuclear states. For some of these pairs (as shown in Table 1), we achieved a factor of $\sim 0.1$ compared to the thermal equilibrium value of $\sim 3 \times 10^{-4}$, representing a gain of nearly three orders of magnitude in signal intensity in a nuclear magnetic resonance experiment. The high nuclear spin polarization achieved in **VO-FP** suggests that photoexcitation of the appended chromophore can be used to significantly enhance the sensitivity of nuclear magnetic resonance (NMR) experiments. In this regard, NMR spectroscopy has already demonstrated that the $^{51}$V in vanadyl porphyrin can be operated as a nuclear qudit coupled to an electron spin ancilla.[31] This notable photoinduced nuclear spin polarization, which has been reported by only a few recent studies in systems involving organic molecular qubits like nitroxides,[16] highlights the urgency for further investigation to fully understand its origin and control it via molecular engineering. Future research needs to explore multi-frequency TREPR measurements or novel time-resolved NMR experiments.

Looking ahead, the proposed system serves as a promising candidate for constructing multi-qubit systems with VO units interconnected by a photo-excitable switch. Trimers, consisting of two VO or VO and Cu qubits linked via FP, could be particularly valuable. The FP photophysical properties could enable it to act as an effective switch for magnetic interactions between the metal qubits. In its ground singlet state, FP is diamagnetic, effectively isolating the qubits and allowing independent single-qubit operations. Upon fast photoexcitation and ISC, FP becomes a spin triplet state, capable of mediating an effective coupling between the qubits, facilitating two-qubit entangling gates.[65] This switching capability can be dynamically controlled using short laser pulses to revert the system to its ground state, providing a mechanism for toggling the coupling on and off. A key requirement for this scheme is maintaining the individual character of each qubit, meaning the exchange interaction $J$ should be rather small compared to the difference in the precession frequencies of the spins, which can be managed through careful bridge engineering.

Overall, our approach lays a foundational building block for the field of light-driven molecular qubits. The synthetic control over the electronic and spin properties of molecular qubits based on transition metals offers vast potential for developing new QIS science materials.

## SUPPORTING INFORMATION

Materials and methods, fsTA/nsTA data at different temperatures and excitation wavelengths, additional CWEPR and TREPR data in toluene and 5CB.

## Notes

The authors declare no competing financial interest.

## ACKNOWLEDGMENT

This work has received funding from the European Union's Horizon Europe research and innovation program under the Marie Skłodowska-Curie Actions Grant Agreement n. 101104276 (PHOTOCODE) and the ERC-Synergy project CASTLE (proj. n. 101071533). Views and opinions expressed are however those of the author(s) only and do not necessarily reflect those of the European Union or the European Commission. Neither the European Union nor the granting authority can be held responsible for them. Angelo C. would like to thank Fondazione Ing. Aldo Gini for a scholarship that made his work abroad possible.

# Supplementary Information for

# Room-Temperature Optical Spin Polarization of an Electron Spin Qudit in a Vanadyl − Free Base Porphyrin Dimer

Alberto Privitera,[1,2*] Alessandro Chiesa,[3] Fabio Santanni,[4] Angelo Carella,[1,5] Davide Ranieri,[4] Andrea Caneschi,[2] Matthew D. Krzyaniak,[1] Ryan M. Young,[1] Michael R. Wasielewski,[1*] Stefano Carretta,[3*] and Roberta Sessoli[4*]

[1] Department of Chemistry, Center for Molecular Quantum Transduction, and Paula M. Trienens Institute for Sustainability and Energy, Northwestern University, 60208-3113, Evanston, IL (USA)

[2] Department of Industrial Engineering, University of Florence & UdR INSTM Firenze, 50139, Firenze, Italy

[3] Department of Mathematical, Physical and Computer Sciences, University of Parma & UdR INSTM, 43124, Parma, Italy.

[4] Department of Chemistry "U. Schiff", University of Florence & UdR INSTM Firenze, 50019, Sesto Fiorentino, Italy

[5] Department of Chemical Sciences, University of Padova, 35134, Padua, Italy

# Table of Contents

# 1. Materials and methods

**Synthesis and characterization.** The free-base monomer, $H_2TrPP$ (**FP**), was purchased from *Porphychem sas* and used as such without further purification. Compounds [VO(TrPP)] (**VO**) and [VOH$_2$(DPP)$_2$] (**VO-FP**) were synthesized according to the procedure in refs.[1-2] Mass spectrometry (MS) and nuclear magnetic resonance spectroscopy on proton ($^1$H-NMR) characterizations are reported here for completeness (Figures S1 – S5). $^1$H-NMR analyses were performed on liquid solutions in CDCl$_3$ (Merck) with a Bruker Advance (400 MHz) instrument and Bruker AV 600 (600 MHz) instruments. Mass spectrometry analyses were carried out using the ThermoFisher LCQ Fleet Ion Trap LC/MS and the Bruker micrOTOF-Q™ III ESI-TOF Mass Spectrometry System equipped with the APCI source.

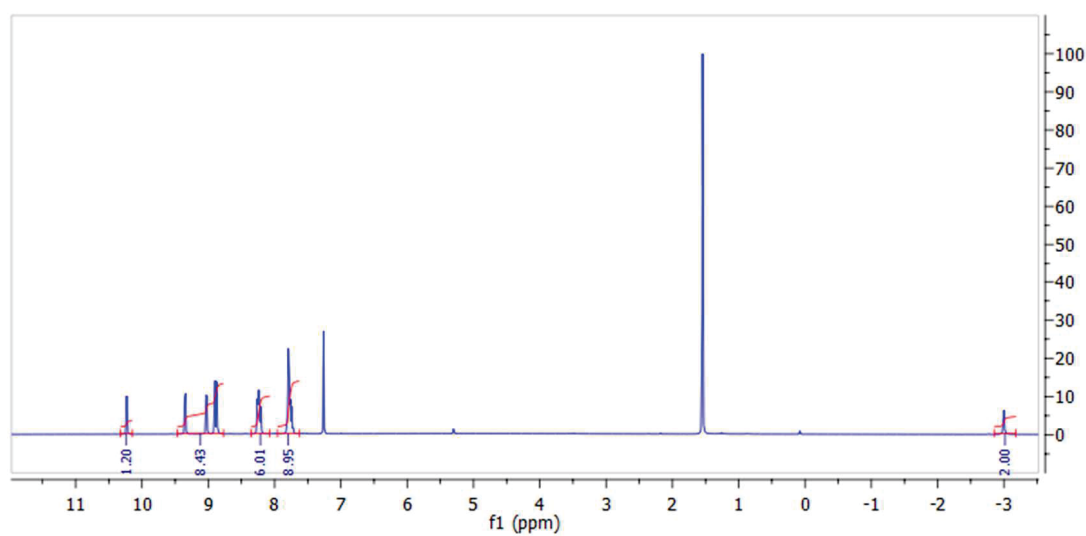

**Figure S1.** $^1$H-NMR spectrum of **FP** in CDCl$_3$ in the range -3.5 – 12 ppm. Non-integrated peaks are ascribed to CH$_2$Cl$_2$ (5.30 ppm), H$_2$O (1. 55 ppm), and silicone grease (0.07 ppm) contaminations.

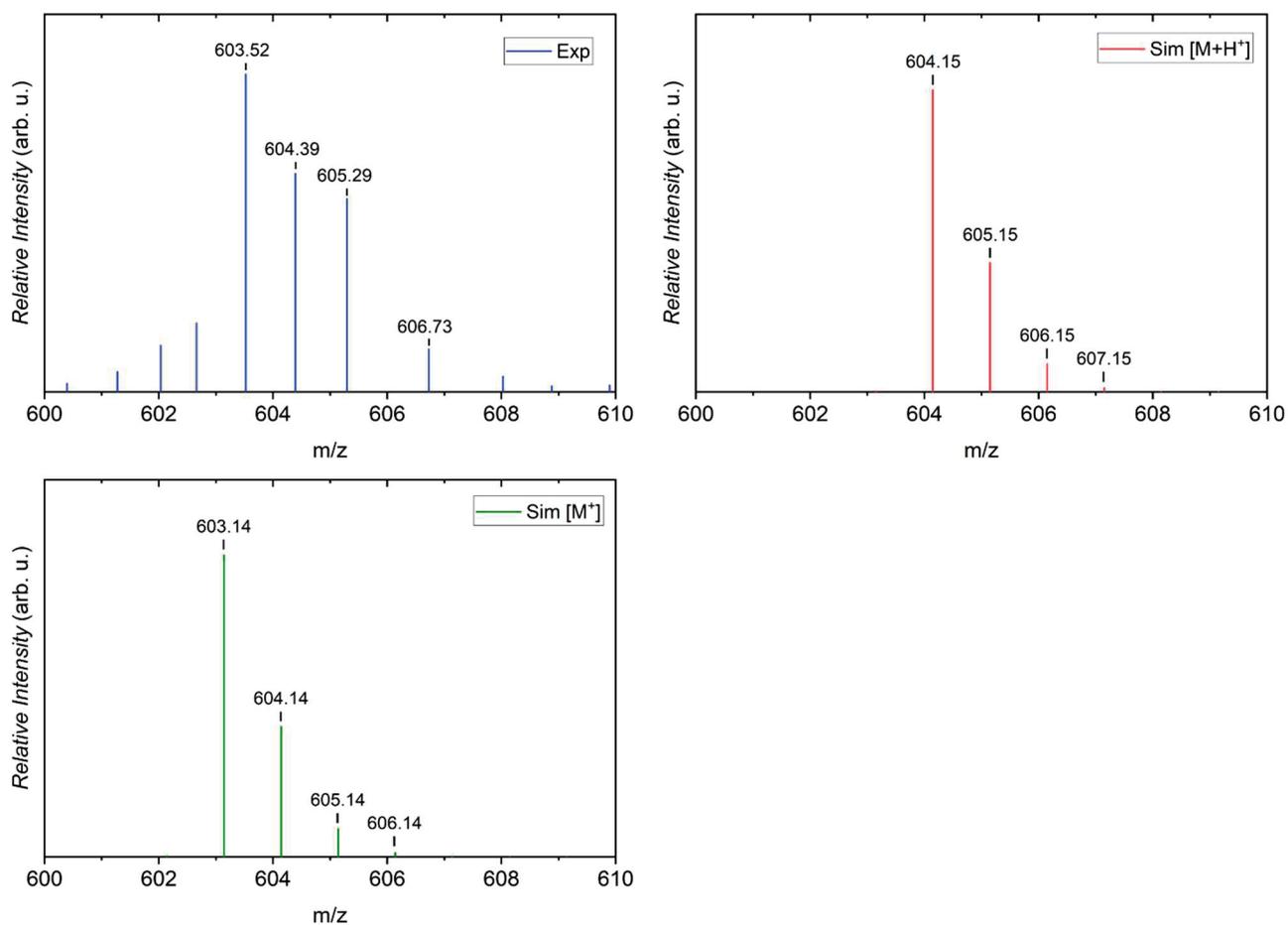

**Figure S2.** Experimental and simulated ESI-MS spectra of **VO**. Small discrepancies in **VO** isotopic pattern can be ascribed to different contributions to the overall spectrum from [M]$^+$ and [M+H]$^+$ species.

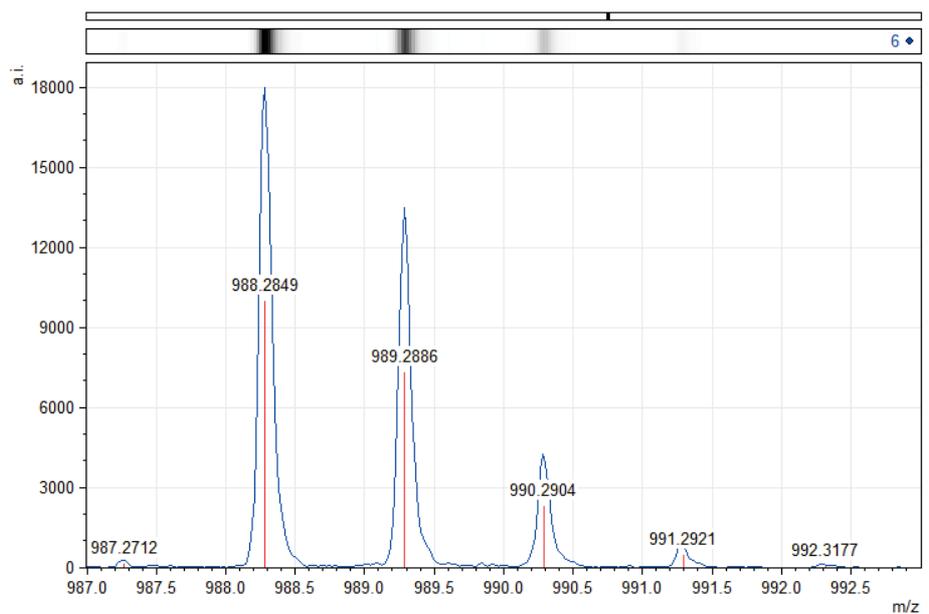
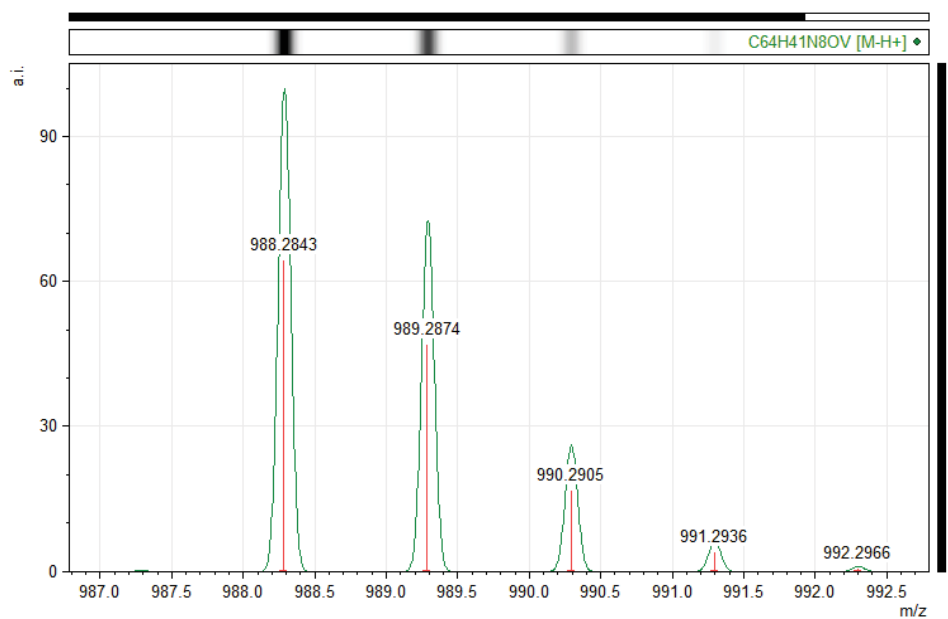

**Figure S3:** Experimental (top, blue line) and simulated (bottom, green line) APCI-MS signals of **VO-FP**.

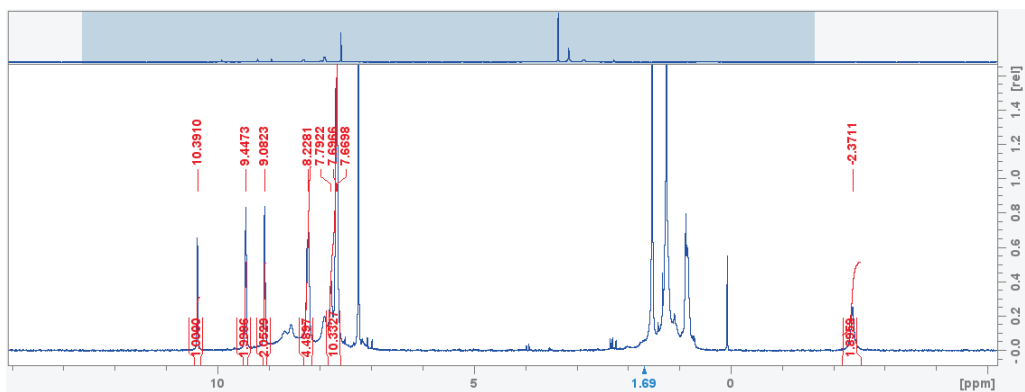

**Figure S4:** $^1$H-NMR spectrum in CDCl$_3$ of **VO-FP**. The signals between 0 and 3 ppm correspond to H$_2$O and grease contamination.

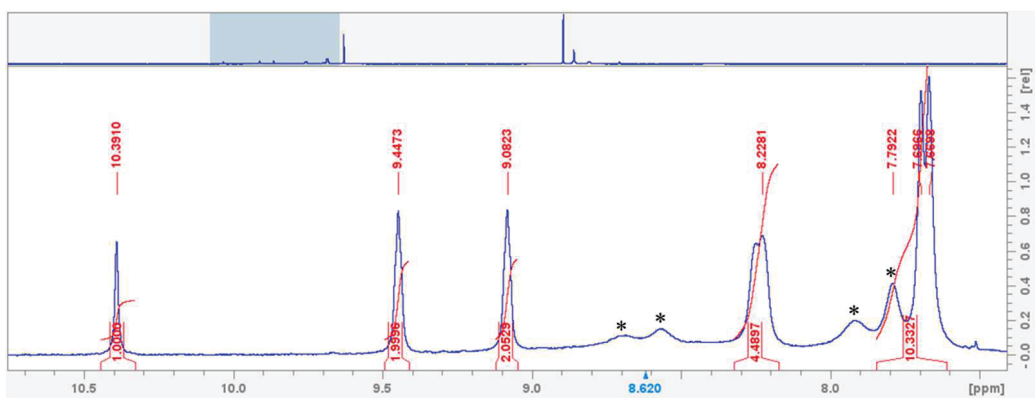

**Figure S5:** $^1$H-NMR signals in CDCl$_3$ of compound **VO-FP** in the aromatic region. Only the signals ascribed to the free-base porphyrin unit were integrated. Broad signals highlighted by an asterisk correspond to $^1$H of the vanadyl porphyrin.

S6

## 2. Transient absorption (TA) spectroscopy

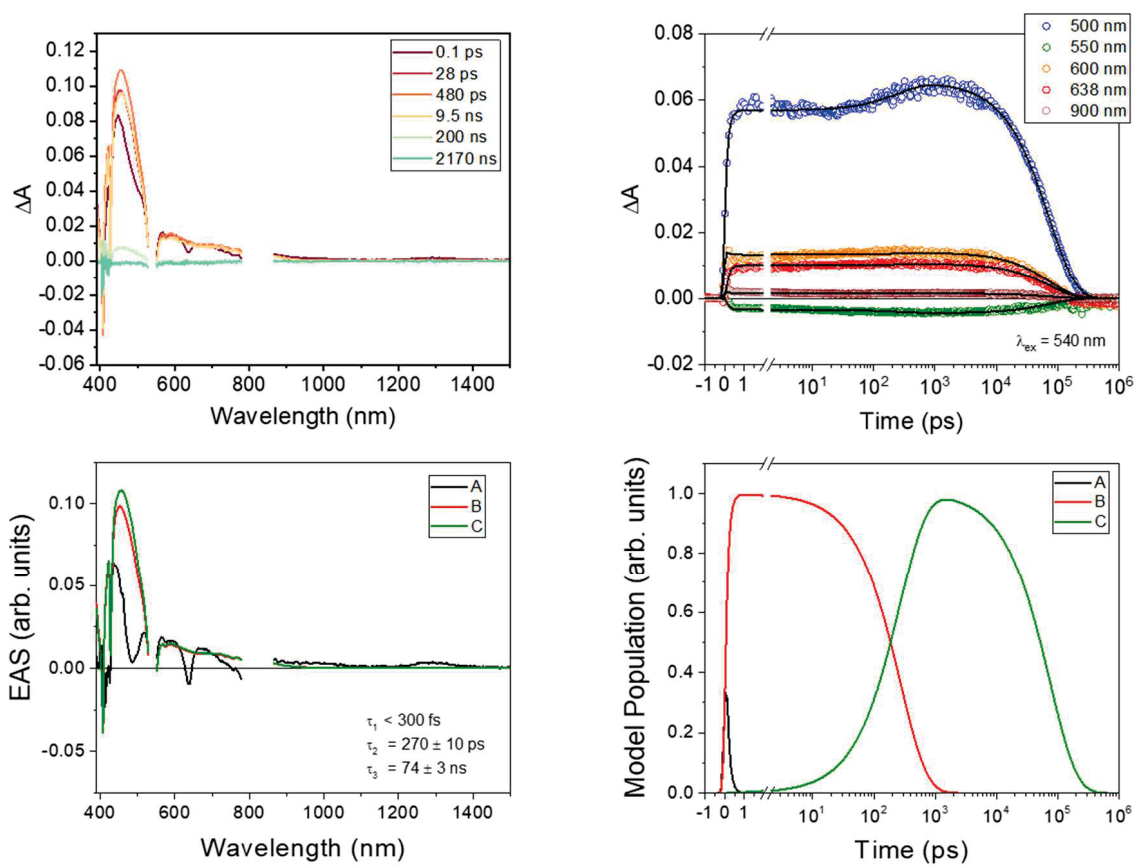

**Figure S6.** (a) Room-temperature fs/nsTA spectra of **VO** in toluene excited at 540 nm and recorded at selected delay times. (b) Selected wavelength kinetic fits, (c) evolution-associated spectra (EAS) and (d) population dynamics obtained by globally fitting the fs/nsTA data. The mechanism assumed to fit the data is A→B→C→ground state, where state A represents the singlet excited state, B is the unrelaxed triplet state, and C is the fully relaxed triplet state.

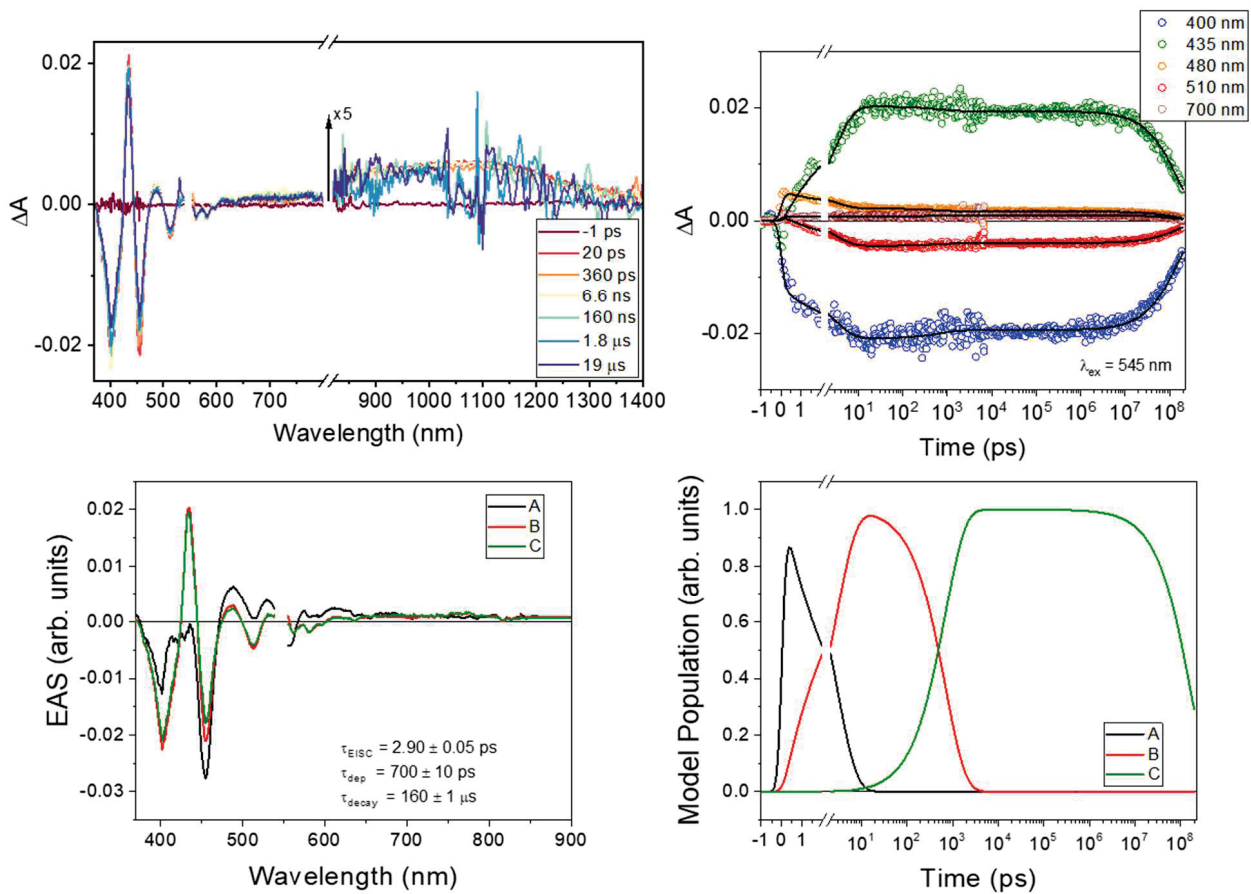

**Figure S7.** (a) Low-temperature (85 K) fs/nsTA spectra of **VO-FP** in butyronitrile excited at 640 nm and recorded at selected delay times. (b) Selected wavelength kinetic fits, (c) evolution-associated spectra (EAS) and (d) population dynamics obtained by globally fitting the fs/nsTA data. The mechanism assumed to fit the data is A→B→C→ground state, where state A represents the singlet excited state, B is the unrelaxed triplet, and C is the geometry-relaxed triplet state. Importantly, no significant photophysical difference is observed between excitation at 640 nm and 545 nm. The only notable difference is a slowdown of the time constants when exciting at 640 nm compared to 545 nm, attributed to slightly faster dynamics at higher photon energies.

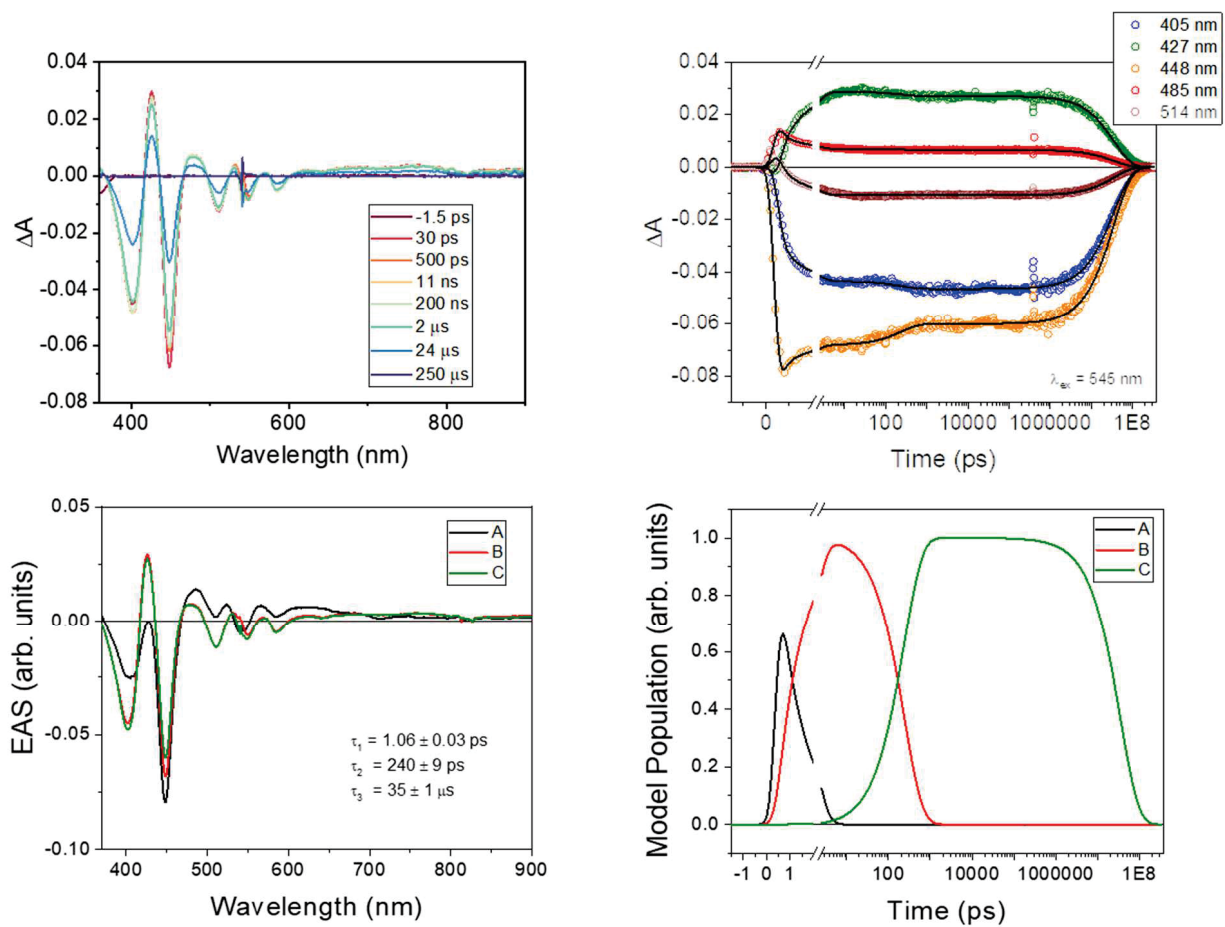

**Figure S8.** (a) Room-temperature fs/nsTA spectra of **VO-FP** in toluene excited at 545 nm and recorded at selected delay times. (b) Selected wavelength kinetic fits, (c) evolution-associated spectra (EAS) and (d) population dynamics obtained by globally fitting the fs/nsTA data. The mechanism assumed to fit the data is A→B→C→ground state, where state A represents the singlet excited state, B is the unrelaxed triplet, and C is the geometry-relaxed triplet state.

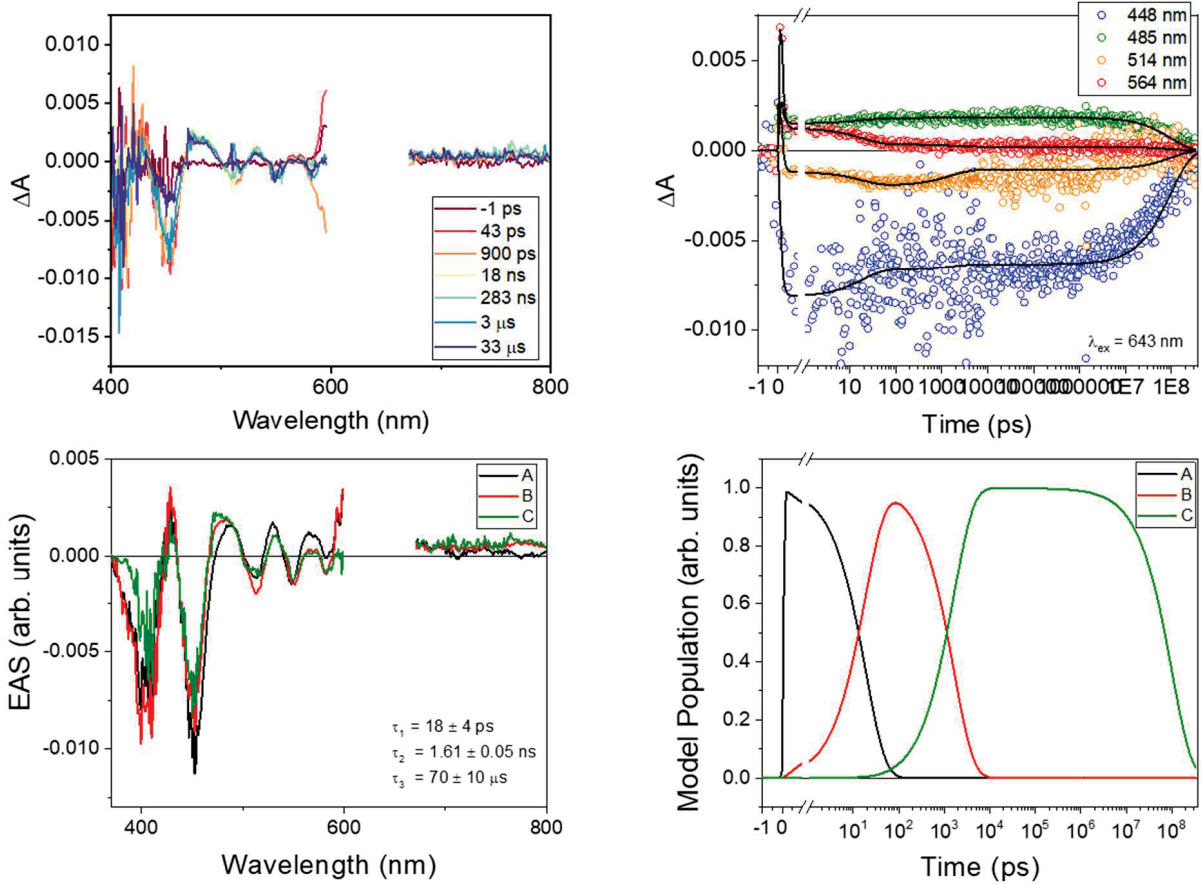

**Figure S9.** (a) Low-temperature (125 K) fs/nsTA spectra of **VO-FP** in toluene excited at 640 nm and recorded at selected delay times. (b) Selected wavelength kinetic fits, (c) evolution-associated spectra (EAS) and (d) population dynamics obtained by globally fitting the fs/nsTA data. The mechanism assumed to fit the data is A→B→C→ground state, where state A represents the singlet excited state, B is the unrelaxed triplet, and C is the geometry-relaxed triplet state. Due to the high laser pump fluence (2 µJ/pulse) used, a relatively intense solvent response at short times is evident. Importantly, no significant photophysical difference is observed between excitation at 640 nm and 545 nm. The only notable difference is a slowdown of the time constants when exciting at 640 nm compared to 545 nm, attributed to slightly faster dynamics at higher photon energies.

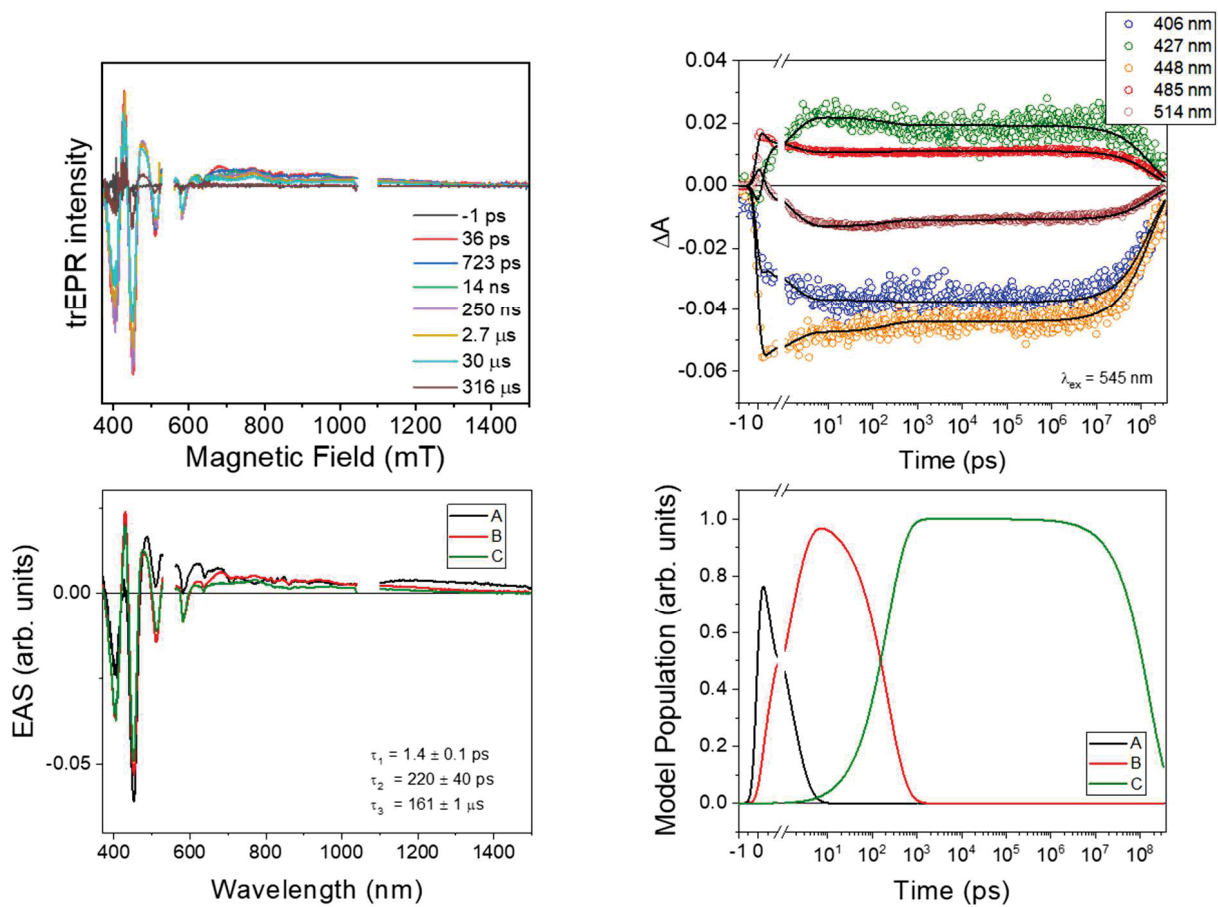

**Figure S10.** (a) Low-temperature (125 K) fs/nsTA spectra of **VO-FP** in toluene excited at 545 nm and recorded at selected delay times. (b) Selected wavelength kinetic fits, (c) evolution-associated spectra (EAS) and (d) population dynamics obtained by globally fitting the fs/nsTA data. The mechanism assumed to fit the data is A→B→C→ground state, where state A represents the singlet excited state, B is the unrelaxed triplet, and C is the geometry-relaxed triplet state.

# 3. Time-resolved Electron Paramagnetic Resonance (TREPR)

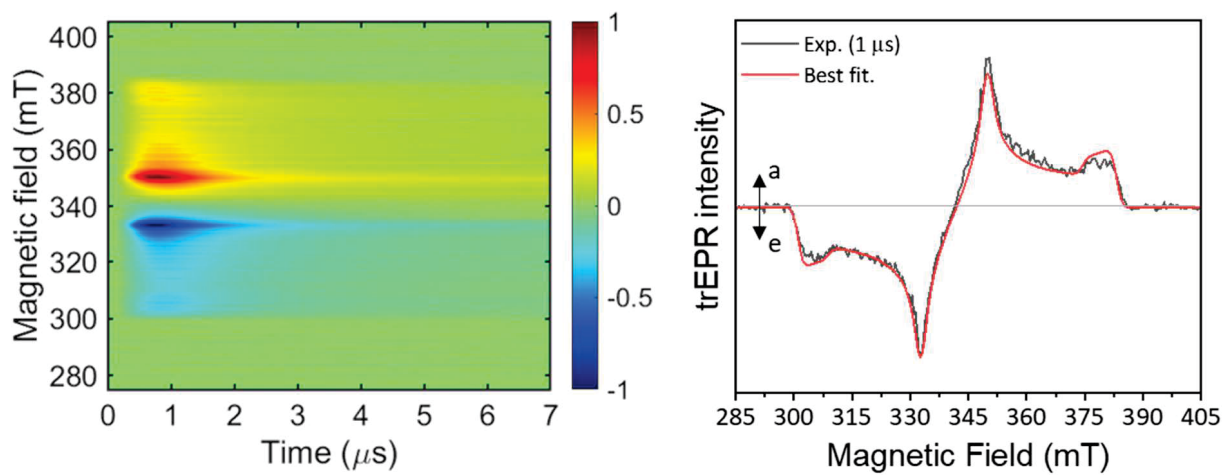

**Figure S11.** (left) Normalised 2D experimental TREPR contour plots of **FP** in toluene acquired at 80 K after a 550 nm laser pulse (7 ns, 2 mJ). Colour legend: red = enhanced absorption, blue = emission, green = baseline. (right) Normalised 1D experimental TREPR spectrum (black line) and best-fit spectral simulation (red line) of $H_2TrPP$ taken at 1 μs after the laser pulse. Arrows legend: a = enhanced absorption, e = emission.

**Table 1.** Best-fit simulations values obtained from the TREPR spectral simulation in Figure S11.

| $g$ | 2.002 |
|---|---|
| $[D, E]$ (MHz) | $[1153, -224] \pm 1$ |
| $[p^x\ p^y\ p^z]$ | $[0.29\ 0.71\ 0] \pm 2$ |
| LW (mT) | $2.5 \pm 0.2$ |

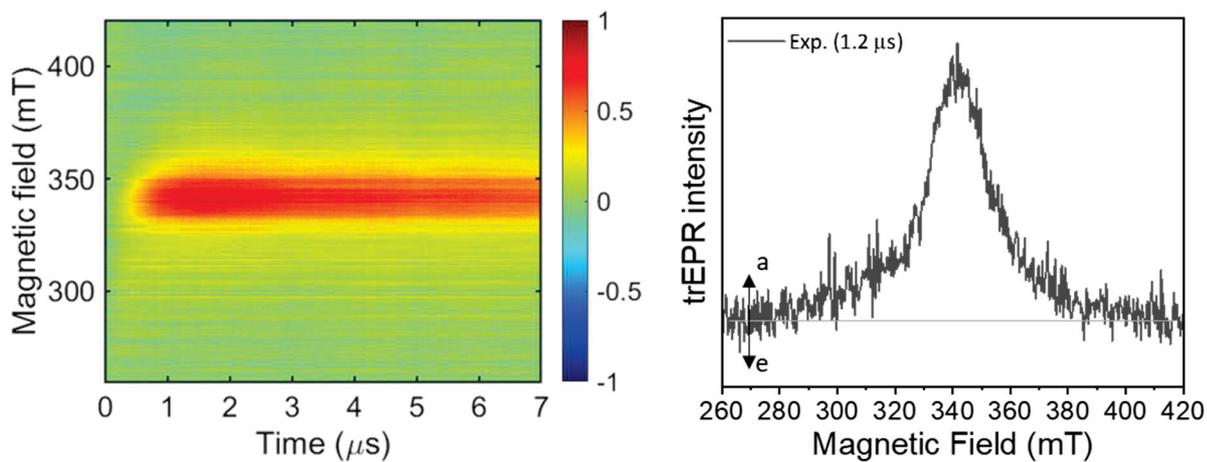

**Figure S12.** (left) Normalized 2D experimental TREPR contour plots of **VO** in toluene acquired at 80 K after a 550 nm laser pulse (7 ns, 2 mJ). Color legend: red = enhanced absorption, blue = emission, green = baseline. (right) Normalized 1D experimental TREPR spectrum of VOTrPP taken at 1.2 μs after the laser pulse. Arrows legend: a = enhanced absorption, e = emission.

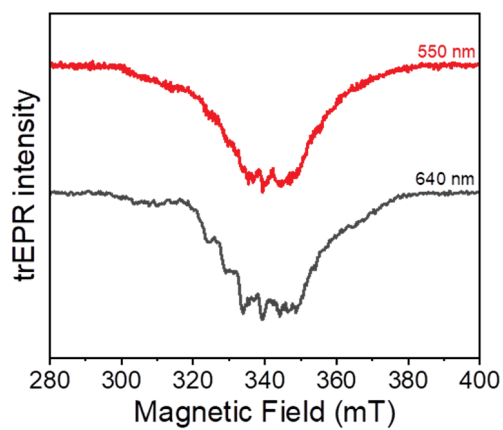

**Figure S13.** Comparison between normalized 1D experimental TREPR spectra of **VO-FP** taken at 1.2 μs after 550nm (red line) and 640nm (black line) laser pulses. The comparison reveals negligible difference between the two spectra, with the exception of a minor line broadening observed in the 550nm spectrum.

## 4. TREPR in 5CB liquid crystal

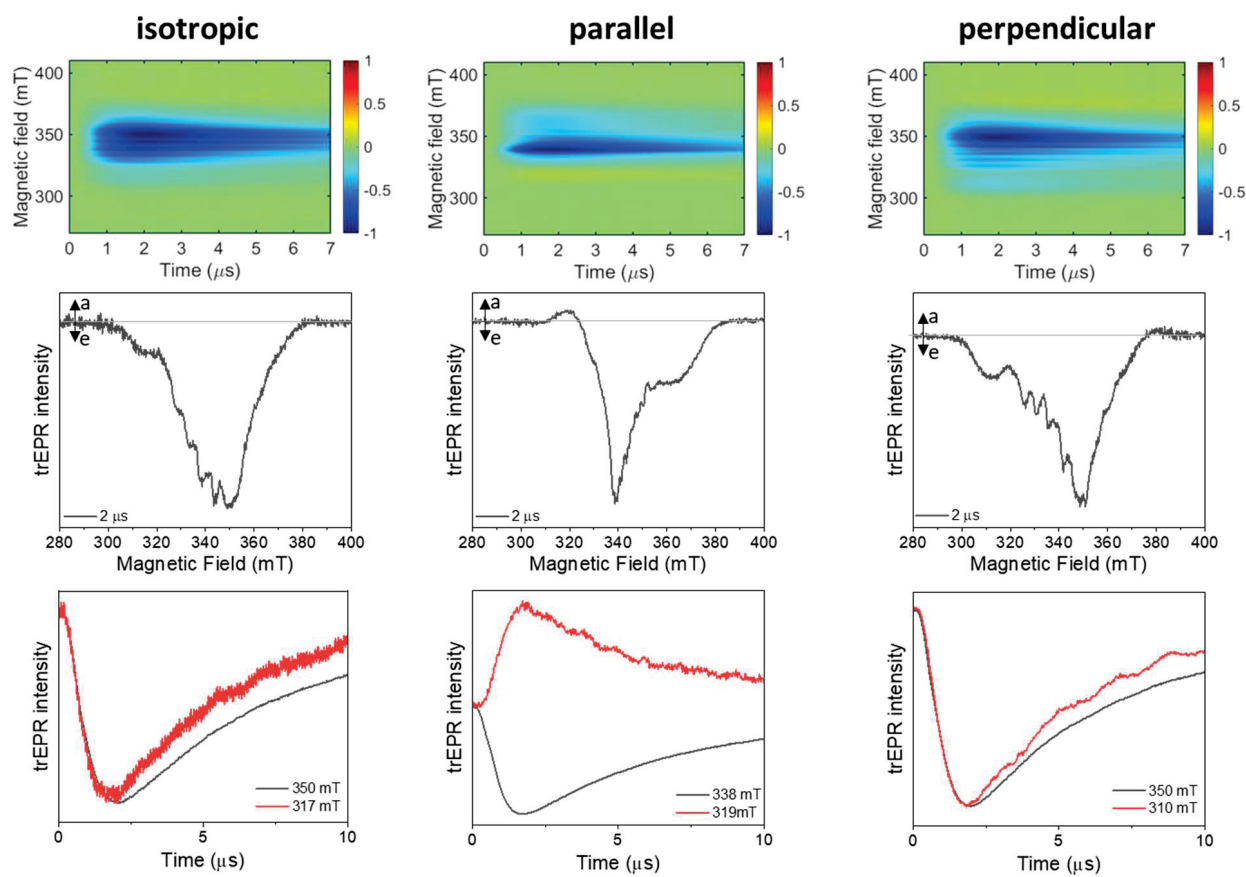

**Figure S14.** (top) Normalized 2D experimental TREPR contour plots of **VO-FP** oriented in the nematic liquid crystal 5CB at 85 K acquired after a 640-nm laser pulse (7 ns, 2 mJ). The long axis of each molecule is aligned at 0° (parallel) and 90° (perpendicular) relative to the applied magnetic field direction. Color legend: red = enhanced absorption, blue = emission, green = baseline. (center) Normalized 1D experimental TREPR spectra taken at 2 µs after the laser pulse. Arrows legend: a = enhanced absorption, e = emission. (bottom) TREPR transients taken at two relevant magnetic field points.

# 5. Continuous-Wave Electron Paramagnetic Resonance (CWEPR)

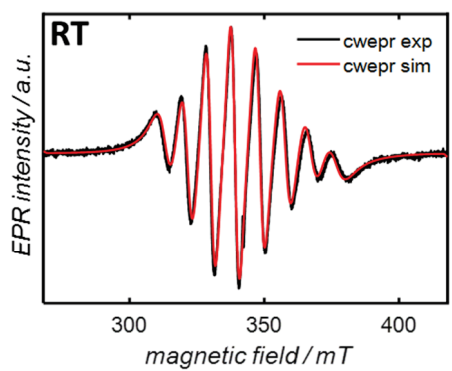

**Chili – slow-motional regime**
Sys.S = 1/2;
Sys.g = [1.985 1.985 1.964];
Sys.Nucs = '51V';
Sys.A =[162 162 475];
Sys.lw = 1.5;
Sys.tcorr = 2.1e-10;  % = 0.21 ns

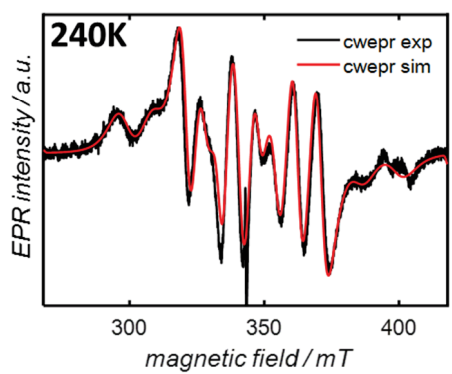

**Chili – slow-motional regime**
Sys.S = 1/2;
Sys.g = [1.985 1.985 1.964];
Sys.Nucs = '51V';
Sys.A =[162 162 475];
Sys.lw = 1.3;
Sys.tcorr = 8e-10;  % = 0.8 ns

**Figure S15.** (left) CWEPR spectra of **VO-FP** (black line) in toluene acquired at room temperature and 240 K. The spectral simulations (red line) were obtained using the routine chili of the Matlab toolbox Easyspin. The simulation parameters used for the simulation are reported on the right.